\documentclass[preprints,advhep,accept,moreauthors,pdftex]{Definitions/mdpi} 

\firstpage{1} 
\pubvolume{1}
\issuenum{1}
\articlenumber{0}
\pubyear{2021}
\copyrightyear{2020}
\datereceived{} 
\dateaccepted{} 
\datepublished{} 
\hreflink{https://doi.org/} 
\pdfoutput=1

\usepackage{xspace}
\newcommand*{\sqs}{\ensuremath{\sqrt{s}}\xspace}
\newcommand*{\Nmpi}{\ensuremath{N_\mathrm{MPI}}\xspace}
\newcommand*{\Nch}{\ensuremath{N_\mathrm{ch}}\xspace}
\newcommand*{\Nfw}{\ensuremath{N_\mathrm{fw}}\xspace}
\newcommand*{\RT}{\ensuremath{R_\mathrm{T}}\xspace}
\newcommand*{\RNC}{\ensuremath{R_\mathrm{NC}}\xspace}
\newcommand*{\SO}{\ensuremath{S_\mathrm{0}}\xspace}
\newcommand*{\pT}{\ensuremath{p_\mathrm{T}}\xspace}

\newcommand*{\GeVc}{\ensuremath{\mathrm{GeV}/c}\xspace}

\newcommand*{\Lc}{\ensuremath{\Lambda_{\rm c}^+}\xspace}
\newcommand*{\Xic}{\ensuremath{\Xi_{\rm c}^{0,+}}\xspace}
\newcommand*{\Sigc}{\ensuremath{\Sigma_{\rm c}^{0,+,++}}\xspace}
\newcommand*{\Omc}{\ensuremath{\Omega_{\rm c}^{0}}\xspace}
\newcommand*{\Dz}{\ensuremath{\rm D^0}\xspace}

\newcommand*{\LcToDz}{\ensuremath{{\Lambda_{\rm c}^+}/{\rm D^0}}\xspace}

\Title{%
Event-activity dependent production of strange and non-strange charmed baryons in the enhanced color-reconnection scheme}

\TitleCitation{Event-activity dependent production of strange and non-strange charmed-baryons}

\Author{Zolt\'an Varga $^{1,2}$*\orcidA{}, Anett Mis\'ak $^{1,3}$, R\'obert V\'ertesi $^{1}$\orcidB{}}

\AuthorNames{Zoltán Varga, Anett Misák, Róbert Vértesi}

\AuthorCitation{Varga, Z.; Misák, A.; Vértesi, R.}
 
\address{%
$^{1}$ \quad Wigner Research Centre for Physics, P.O. Box 49, H-1525 Budapest, Hungary\\
$^{2}$ \quad Budapest University of Technology and Economics, M\H{u}egyetem rkp. 3., H-1111 Budapest, Hungary\\
$^{3}$ \quad Eötvös Loránd University, Budapest, Hungary}

\corres{Correspondence: varga.zoltan@wigner.hu}

\abstract{We investigated the production of charmed baryons with different isospin and strangeness content, compared to both charmed \Dz mesons and to the \Lc baryon in proton--proton collisions at LHC energies. We used the PYTHIA 8 Monte Carlo event generator with color-reconnection beyond leading color approximation and proposed methods based on event-activity classifiers to probe the source of the charm baryon enhancement. We conclude that in the considered model class, the isospin of the charmed baryon state has a strong impact on the enhancement pattern. Using the observables we propose, upcoming high-precision experimental data will be able to differentiate between mechanisms of strangeness and charm enhancement.}


\begin{document}
\end{paracol}

\section{Introduction}\label{Intro}

Perturbative quantum-chromodynamics (pQCD) calculations~\cite{Kniehl:2012ti,Cacciari:2012ny} have been successful in describing the production of heavy-flavor mesons at several energies at the LHC~\cite{ALICE:2021mgk,LHCb:2017yua,ALICE:2017olh}.
These descriptions usually rely on the factorization approach, in which the production cross sections of heavy-flavor hadrons in hadronic collisions are calculated as a convolution of the parton density functions (PDF) of the colliding hadrons, the cross section of the hard scattering processes that produce the heavy quarks, and the fragmentation function that describes the probability of the formation of the given heavy-flavor hadron from the heavy quark~\cite{Collins:1989gx}.
Since the three components are treated as independent terms in this approach, fragmentation functions are assumed to be universal and are often determined from ${\rm e}^{-}{\rm e}^{+}$ (or ${\rm e}^{-}{\rm p}$) collisions where PDF plays no (or less important) role~\cite{Braaten:1994bz}. 
Recent experimental results by ALICE, CMS and LHCb on the production of charmed baryons, however, do no not support the factorization approach with universal fragmentation: ratios of charmed-baryon yields to that of charmed mesons show a transverse-momentum (\pT) dependent enhancement compared to ${\rm e}^{-}{\rm e}^{+}$ results, at both central and forward rapidity, that is mostly concentrated in the soft to semi-soft ($\pT \lesssim 8$ GeV/$c$) momentum range~\cite{ALICE:2017thy,CMS:2019uws,ALICE:2020wfu,ALICE:2021bli,LHCb:2018weo}.
%

Several scenarios have been proposed to explain the enhancement. Coalescence models include the quark combination mechanism (QCM), based on statistical weights and equal quark-velocity~\cite{Song:2018tpv}, and the Catania model, which implements the recombination of charm quarks with light quarks in the hot QCD matter~\cite{Plumari:2017ntm}. A statistical hadronization model (SHM)~\cite{He:2019tik} considers the feed-down from several higher-mass charm states beyond those listed by the particle data group (PDG)~\cite{ParticleDataGroup:2020ssz}. 
Another model class considers color-reconnection with string formation beyond leading color approximation (CR-BLC)~\cite{Christiansen:2015yqa}, by introducing color junctions into the PYTHIA 8~\cite{Sjostrand:2014zea} multiple-parton interactions (MPI) framework with color reconnection (CR)~\cite{Sjostrand:2017cdm}. 
While all of these models describe the enhanced \LcToDz ratios quantitatively~\cite{ALICE:2020wfu}, predictions for the  measurements of \Sigc, \Xic and \Omc ratios to charmed mesons and baryons are sometimes not adequate or only able to provide a qualitative match~\cite{ALICE:2021bli,ALICE:2022cop,ALICE:2021psx,ALICE:2021rzj}. The strangeness content ($s$) and isospin ($I$) of the given charmed baryon states, as well as feed-down contributions, may play an important role in the differences.

In our recent work~\cite{Varga:2021jzb} we argued that the \LcToDz enhancement with respect to different event-activity classifiers provide sensitive probes that can access the source of the enhanced charmed-baryon production, and thus differentiate between the above scenarios.
In the enhanced-CR model~\cite{Christiansen:2015yqa} the excess is linked to the MPI, which is strongly correlated with the underlying-event (UE) activity~\cite{Martin:2016igp}. We used this model to conduct detailed studies for the \LcToDz ratios in proton-proton collisions at \sqs=13 TeV. 
We saw that \LcToDz enhancement strongly depends on the final-state hadron multiplicity both in the central and the forward rapidity region.
This behavior is supported by recent ALICE measurements that had been since finalized~\cite{ALICE:2021npz}, hinting that the multiplicity taken from forward and central regions show the same enhancement qualities, therefore attesting to the usability of the enhanced-CR scenario. We also proposed to observe the \LcToDz ratio in event classes that are determined based on the activity within the underlying event, as well as the multiplicity of the jet caused by the leading process, and concluded that the charmed-baryon enhancement can be observed to depend on the UE activity but not on the activity inside the jet region.
%

In the current work we extend our study from \Lc(qqc, $I=0$) to the charmed baryon states \Sigc(qqc, $I=1$), \Xic(qsc) and \Omc(ssc), to pin down the relative contributions of strangeness and charm in the baryon enhancement, and also address the role of the isospin. While we phrase our observations in terms of MPI using the enhanced-CR scenario~\cite{Christiansen:2015yqa}, we provide predictions for observables that are accessible for the experiment already in the LHC Run3 data-taking phase.

\section{Analysis Method}
We used the PYTHIA 8.303~\cite{Sjostrand:2014zea} general-purpose Monte-Carlo event generator with the Monash tune and the SoftQCD settings~\cite{Skands:2014pea}, with the CR-BLC model mode 2~\cite{Christiansen:2015yqa}, to generate proton-proton (pp) collision events at $\sqs=13$ TeV center-of-mass energy. 
PYTHIA models a basic hard scattering process with leading-order pQCD calculations, combined with initial- and final-state radiations as well as MPI at the partonic level. 
The hadronic final state is produced using Lund string fragmentation, and then secondary decays and rescattering between hadrons are computed, forming the final state of the collision.
The Monash 2013 tune is mainly focused on describing the minimum-bias and underlying event (UE) distributions accurately. 
We simulated 1.2 billion events with the following settings.

We quantified the event activity at central pseudorapidity using the event multiplicity \Nch, defined as the number of all charged final state particles with a minimum transverse momentum $\pT = 0.15~\GeVc$ within the pseudorapidity window $|\eta| < 1$ and full azimuth coverage.
We also used the number of final state particles in the $2 < |\eta| < 5$, \Nfw, to classify events based on their activity in the forward pseudorapidity range.

In events where the leading (highest transverse-momentum) charged-particle track within $|\eta|<1$ fulfilled the condition of $p_\mathrm{T}^{\mathrm{leading}}>5$~GeV/$c$ (also called the trigger hadron), we utilized the event-classifier $R_\mathrm{T} \equiv N_\mathrm{ch}^{\mathrm{transverse}}/\langle N_\mathrm{ch}^{\mathrm{transverse}}\rangle$~\cite{Martin:2016igp}, where the transverse multiplicity $N_\mathrm{ch}^{\mathrm{transverse}}$ is defined as the number of charged-particle tracks in the transverse region $\frac{\pi}{3} < \Delta\phi < \frac{2\pi}{3}$, with $\phi$ being the azimuth angle of the track.
Note that in PYTHIA, which describes the event in terms of MPI, the \RT event-classifier is strongly correlated with the number of MPI~\cite{Martin:2016igp}.

Analogously to the above, we used the near-side cone event-classifier $R_\mathrm{NC} \equiv N_\mathrm{ch}^{\mathrm{near\text{-}side\; cone}}/\langle N_\mathrm{ch}^{\mathrm{near\text{-}side\; cone}}\rangle$~\cite{Varga:2021jzb}, which is the self-normalized multiplicity in a narrow cone around the trigger particle with radius $\sqrt{\Delta\phi^2+\Delta\eta^2}<0.5$. Since the activity inside this narrow cone is mainly dominated by the leading particle and the jet fragments, the \RNC quantity is inherently linked to the activity of the jet region, so the \RNC correlates well with the jet activity.

To quantify the jettiness and underlying event activity of events without a high-\pT trigger particle, we used the transverse spherocity \SO~\cite{Ortiz:2015ttf}, defined as
\begin{eqnarray}
\SO \equiv \frac{\pi}{4} \min\limits_{\bf \hat{n}} \left( \frac{\sum_i \left| {\bf p}_{{\rm T},i} \times {\bf \hat{n}} \right| }{\sum_i p_{{\rm T},i}} \right)\ , 
\end{eqnarray} 
where the $i$ index runs over all accepted particles and ${\bf \hat{n}}$ is any unit vector in the azimuth plane. 
Spherocity is defined with a number between 0 and 1. Isotropic events correspond to $\SO\rightarrow 1$, while $\SO\rightarrow 0$ for jetty events with strongly collimated particles.
The event-activity classes were determined as described in Table 1 of Ref.~\cite{Varga:2021jzb}, in a way that each of them contains a similar amount of events.


While \Nmpi is not a physically observable quantity, it is closely related to the observed charmed-baryon enhancement in the model class under investigation~\cite{Varga:2021jzb}. Therefore we used \Nmpi as the most handy quantity to represent the relation of the enhancement to the final-state event activity. Later on we also propose easily accessible physical observables that are powerful in distinguishing between different model scenarios and can be directly used in future data comparison.

In each event we selected charm hadrons in the central rapidity window $|y| < 0.5$.
We analysed the charmed baryons
  $ \Lambda ^ {+} _ {c} $, 
  \Sigc, 
  \Xic, $ \Omega ^ {0} _ {c} $ and $ \Omega^ {*0} _ {c} $ (commonly referred to as \Omc), as well as the charmed \Dz meson.
We excluded feed-down from beauty hadrons. In case of \Lc 
a substantial contribution comes from decays of other charm-baryon (predominantly the \Sigc) states. Based on event-generator information we evaluated \Lc contributions from direct hadronization, i.e. separately from those that stem from the decay of \Sigc baryons.
Since the meson-to-baryon and baryon-to-baryon ratios are sensitive to differences in fragmentation mechanisms without sensitivity to the heavy-flavor production cross section, we used the ratios of the different charmed-baryon to the \Dz meson as well as to the \Lc, in function of \pT and different event activity classes.

\section{Results and Discussion}
\begin{figure}[ht!]
\begin{center}
\includegraphics[width=0.33\columnwidth]{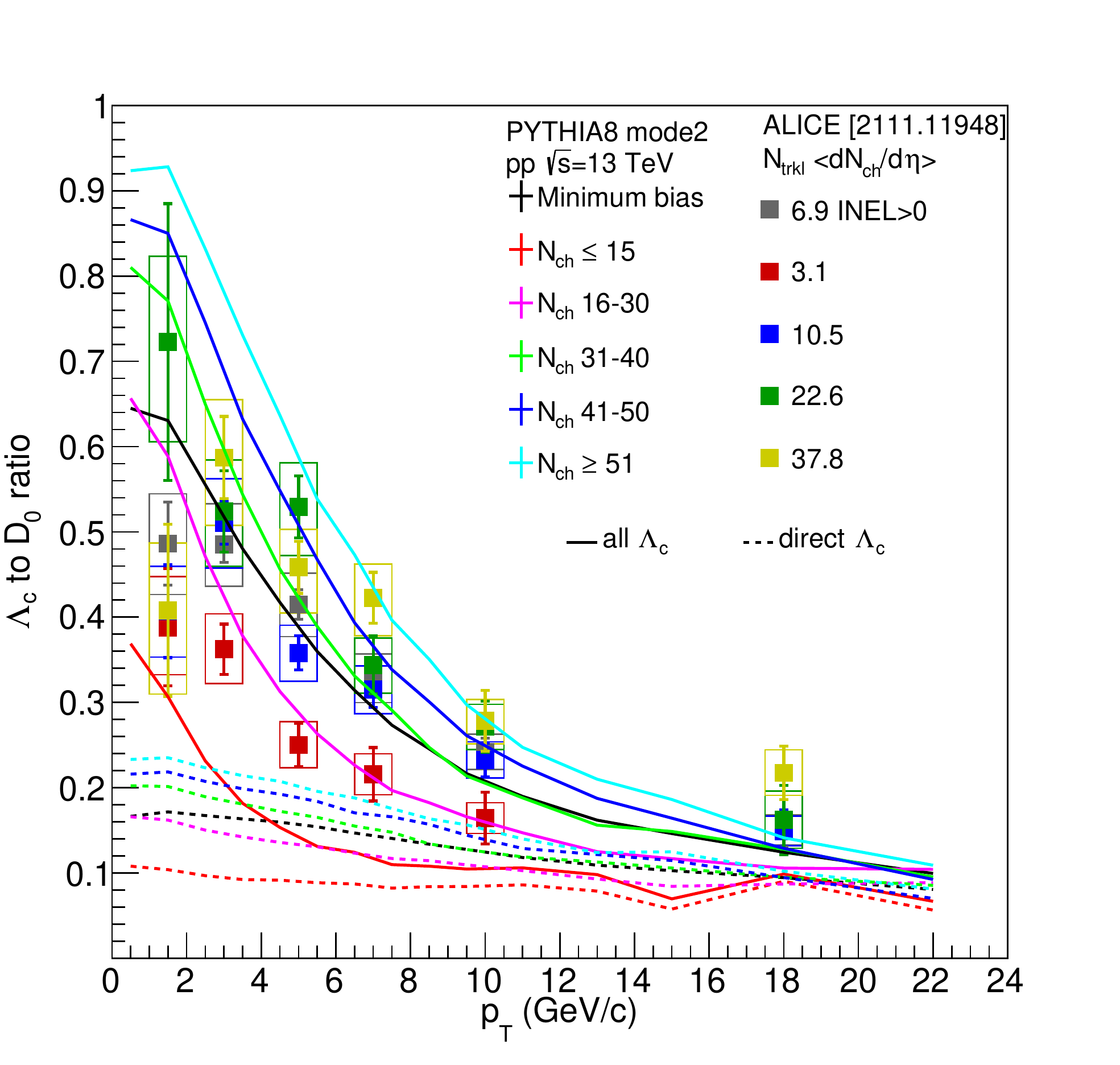}%
\includegraphics[width=0.33\columnwidth]{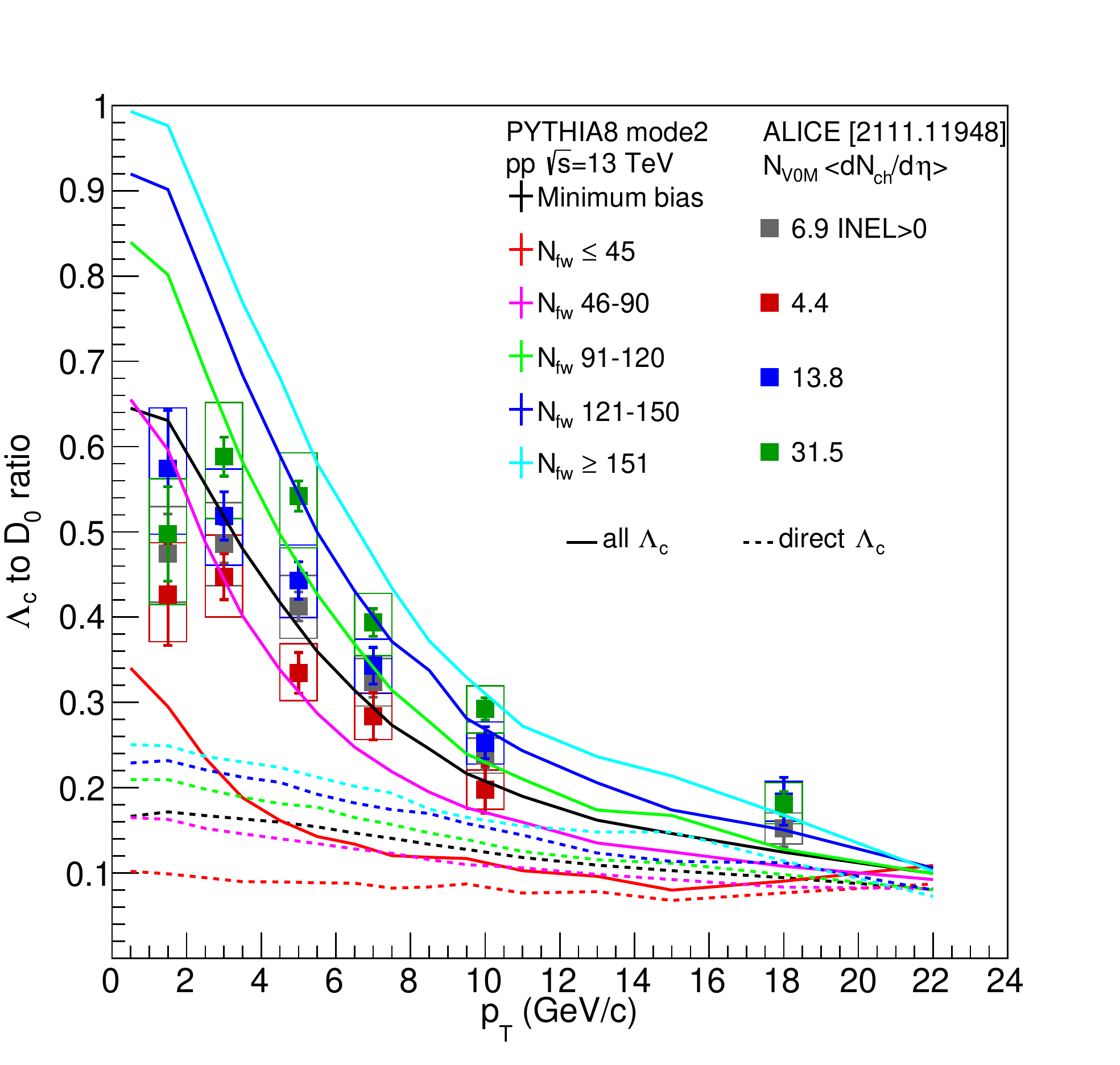}%
\includegraphics[width=0.33\columnwidth]{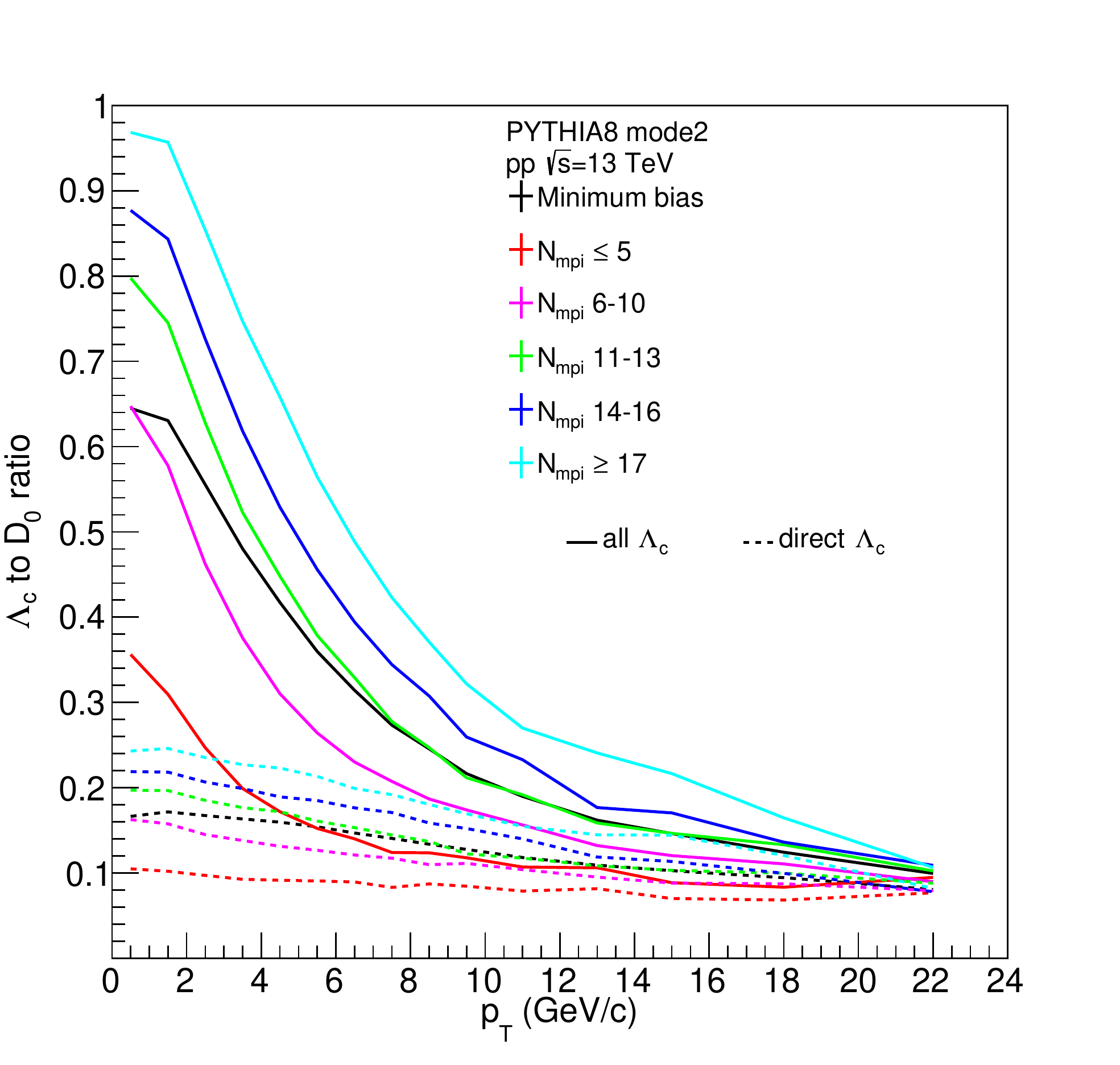}%
\caption{\LcToDz ratio from PYTHIA 8 simulations with enhanced CR as a function of \pT, for MB events as well as for different mid-rapidity (\Nch, left) and forward-rapidity (\Nfw, center) charged-hadron multiplicity as well as multiparton-interaction (\Nmpi, right) classes, shown as solid lines, compared to data from ALICE~\cite{ALICE:2021npz}. The contribution of direct \Lc production is shown separately as dashed lines. (Center:) $\Sigc/\Dz$ ratios for MB events and in different MPI classes. (Right:) $\Sigc/\Lc$ ratios for MB events and in different MPI classes.}
\label{fig:LcVsD0nch}
\end{center}
\end{figure}

In Fig.~\ref{fig:LcVsD0nch}  we show the \LcToDz ratios in different \Nch, \Nfw and \Nmpi bins (left, center and right panels, respectively) for PYTHIA 8 with CR-BLC. In all three panels we separately show the contribution from \Lc baryons directly produced in the hadronization, as dashed lines. We compare the simulations to recently published central and forward data from Ref.~\cite{ALICE:2021npz} (left and center panels).
The calculations from PYTHIA 8 with CR-BLC reproduce data trends in which the enhancement of the \LcToDz ratio strongly correlates with the event multiplicities. It should be noted that the multiplicity classes in the model calculations do not correspond exactly to the ones from Ref.~\cite{ALICE:2021npz}, and in case of forward rapidity the pseudorapidity definition also differs. In the ALICE analysis the multiplicity in the forward rapidity region has been estimated from the percentile distribution of the sum of signal amplitudes in the V0A and V0C scintillators, covering the pseudorapidity regions $2.8 < |\eta| < 5.1$ and $-3.7 < |\eta| < -1.7$, respectively.
As mentioned in Sec.~\ref{Intro}, in the scenario described by PYTHIA 8 with CR-BLC, the low-\pT \LcToDz excess is bound to the underlying-event activity, which is well represented by \Nmpi in the model. While in this case the very same dependence on \Nmpi can be observed similarly to the case of the \Nch and \Nfw classes, classification based on observables linked to \Nmpi can better differentiate between specific scenarios~\cite{Varga:2021jzb}.

Besides \Lc baryons originating directly from the hadronization, there is a significant contribution from decay products of other charm hadrons (predominantly the \Sigc states), that even exceeds direct production at low \pT~\cite{ALICE:2021rzj}. These two charm baryonic states differ in their isospin, which may influence the production of these particles depending on the hadronization mechanism. Both contributions are ordered with event activity. However, secondary \Lc production from \Sigc decays dominates the low-\pT range and diminishes toward higher \pT, the contribution of direct \Lc is relatively flat in \pT, and thus dominates the high-\pT range.
\begin{figure}[ht!]
\begin{center}
\includegraphics[width=0.33\columnwidth]{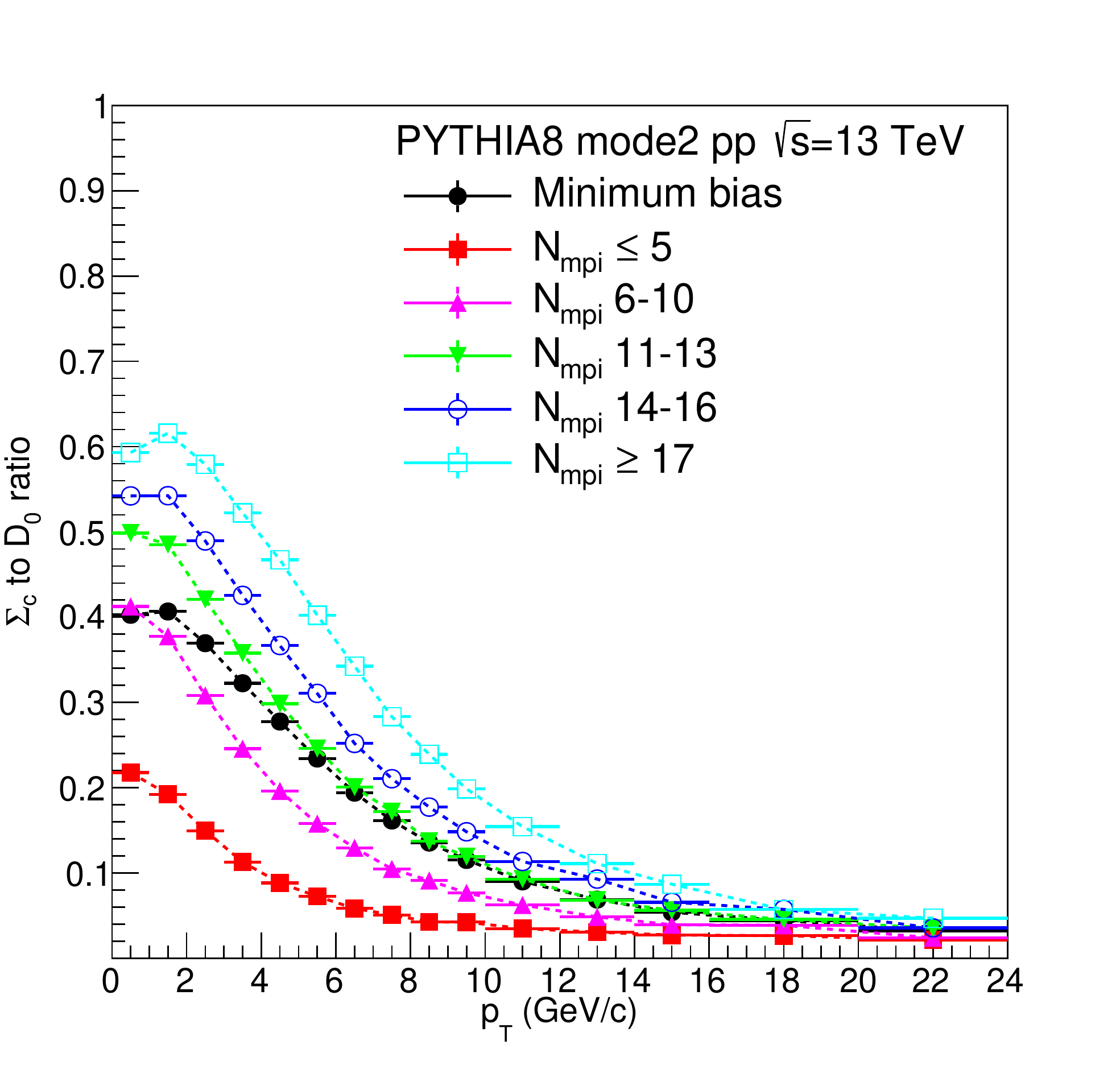}%
\includegraphics[width=0.33\columnwidth]{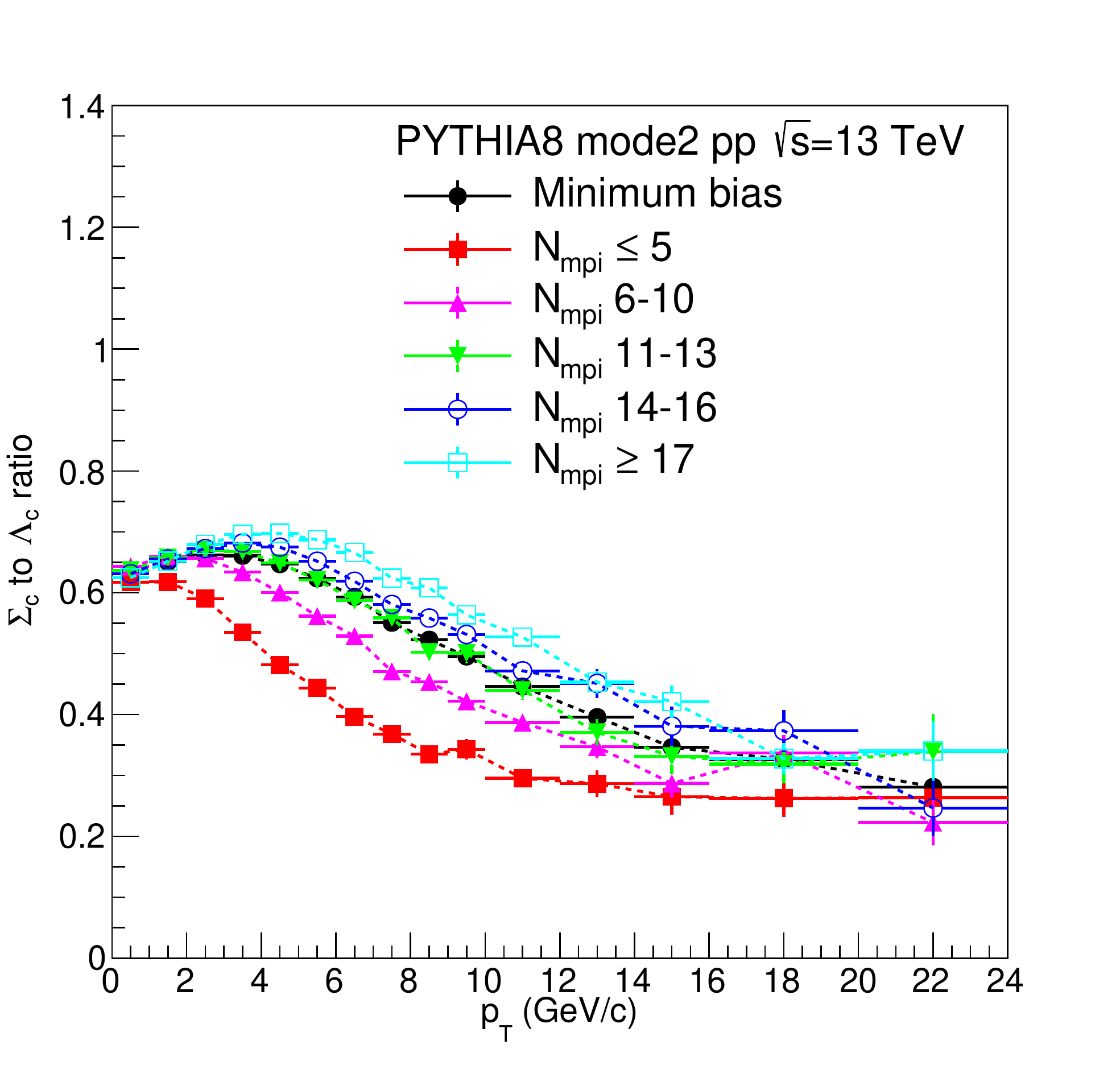}%
\caption{$\Sigc/\Dz$ (left) and $\Sigc/\Lc$ (right) ratios from PYTHIA 8 simulations with enhanced CR as a function of \pT, for MB events as well as for different multiparton-interaction (\Nmpi) classes.}
\label{fig:SigcVsD0nch}
\end{center}
\end{figure}   
In Fig.~\ref{fig:SigcVsD0nch} (left) we show the ratios of $\Sigc$ yields both to the \Dz meson and to \Lc, in different \Nmpi classes. The $\Sigc/\Dz$ ratio shows the ordering by \Nmpi (and \Nch) the very same way as it is present in the \Lc from \Sigc decays. The differences in trends between \Lc and \Sigc are highlighted in the baryon-to-baryon ratios of Fig.~\ref{fig:SigcVsD0nch} (right). Since this ratio is also ordered by \Nmpi, we can exclude the effect of decay kinematics.

\begin{figure}[ht!]
\includegraphics[width=0.33\columnwidth]{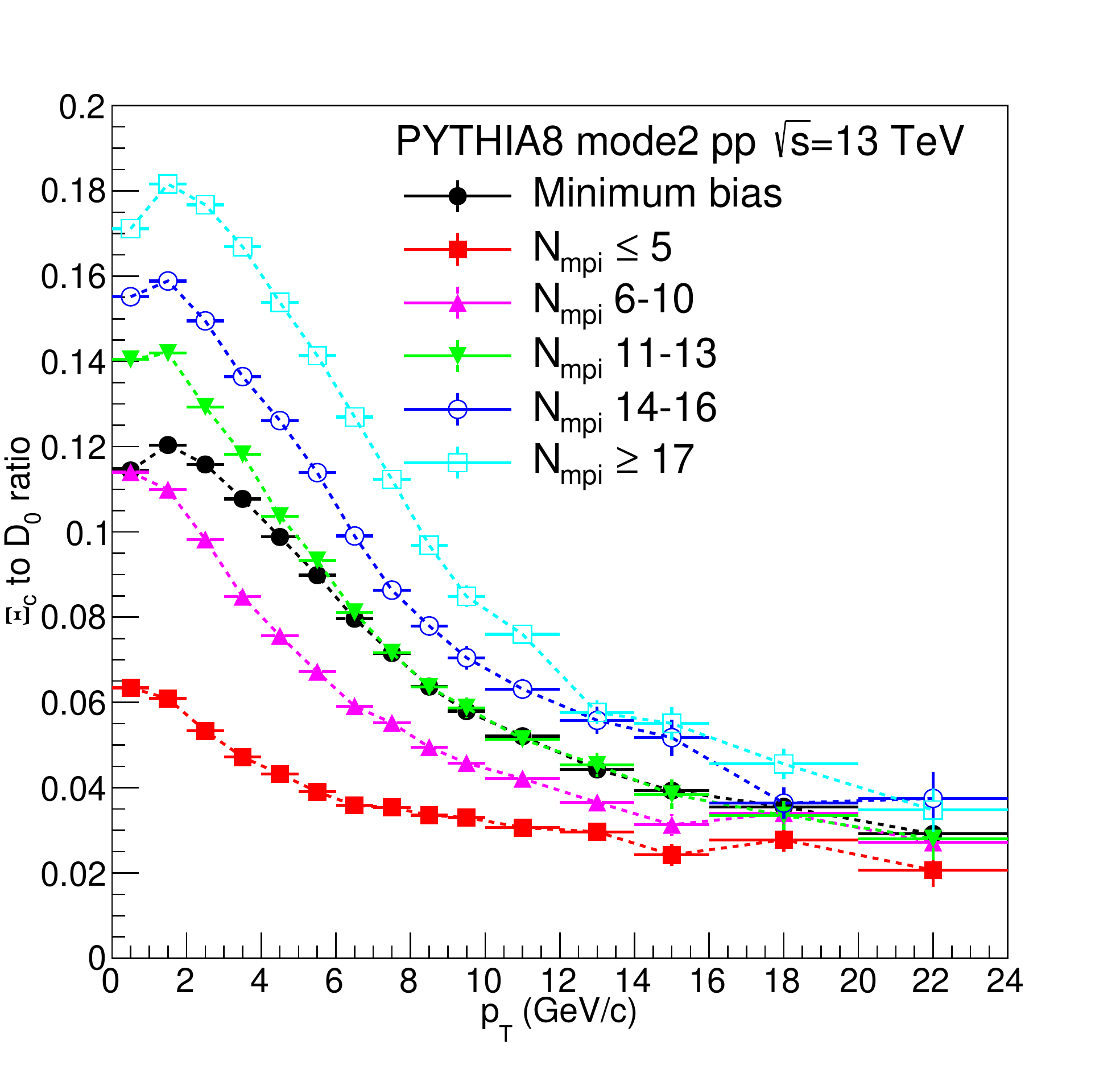}%
\includegraphics[width=0.33\columnwidth]{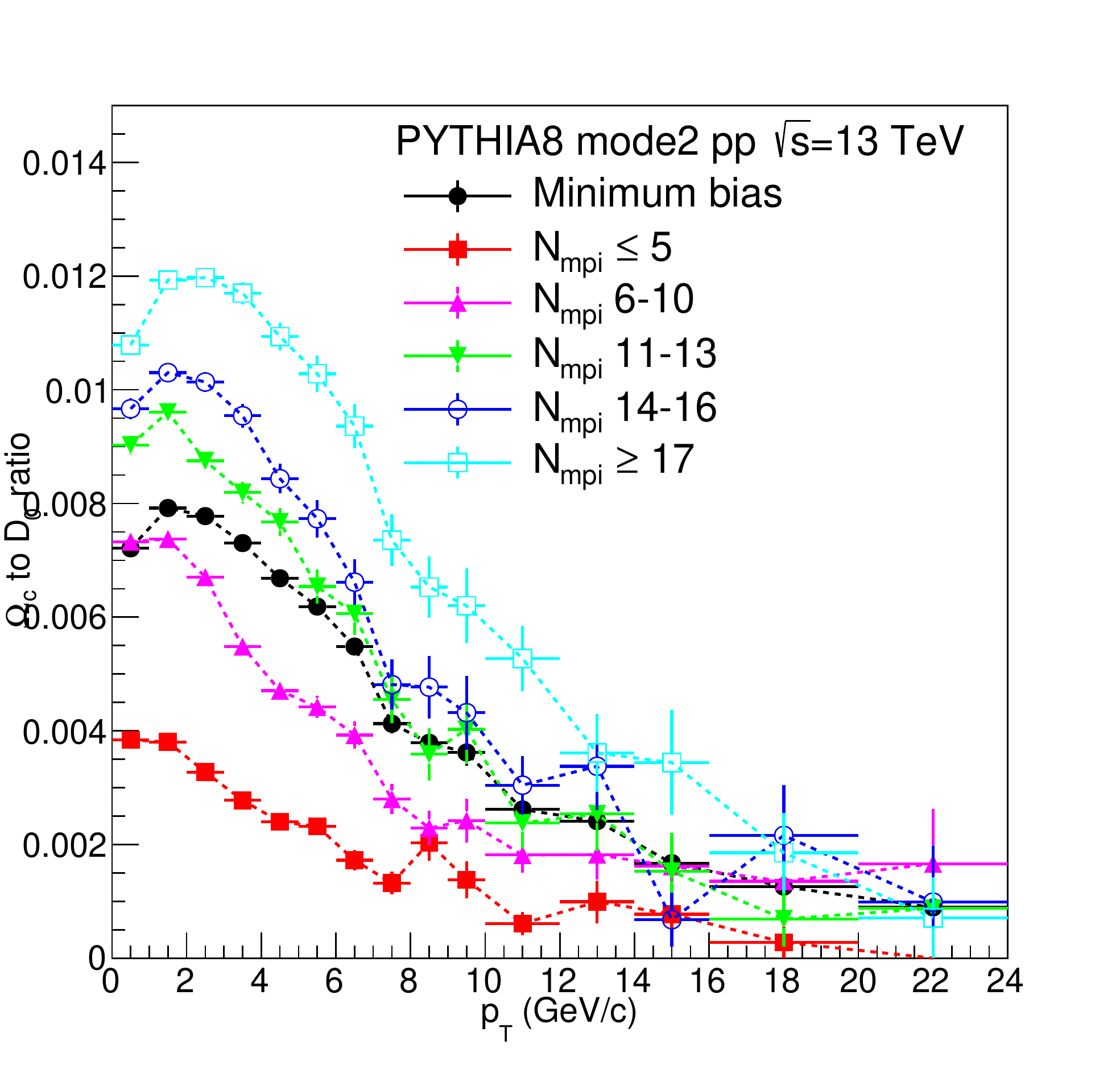}
\includegraphics[width=0.33\columnwidth]{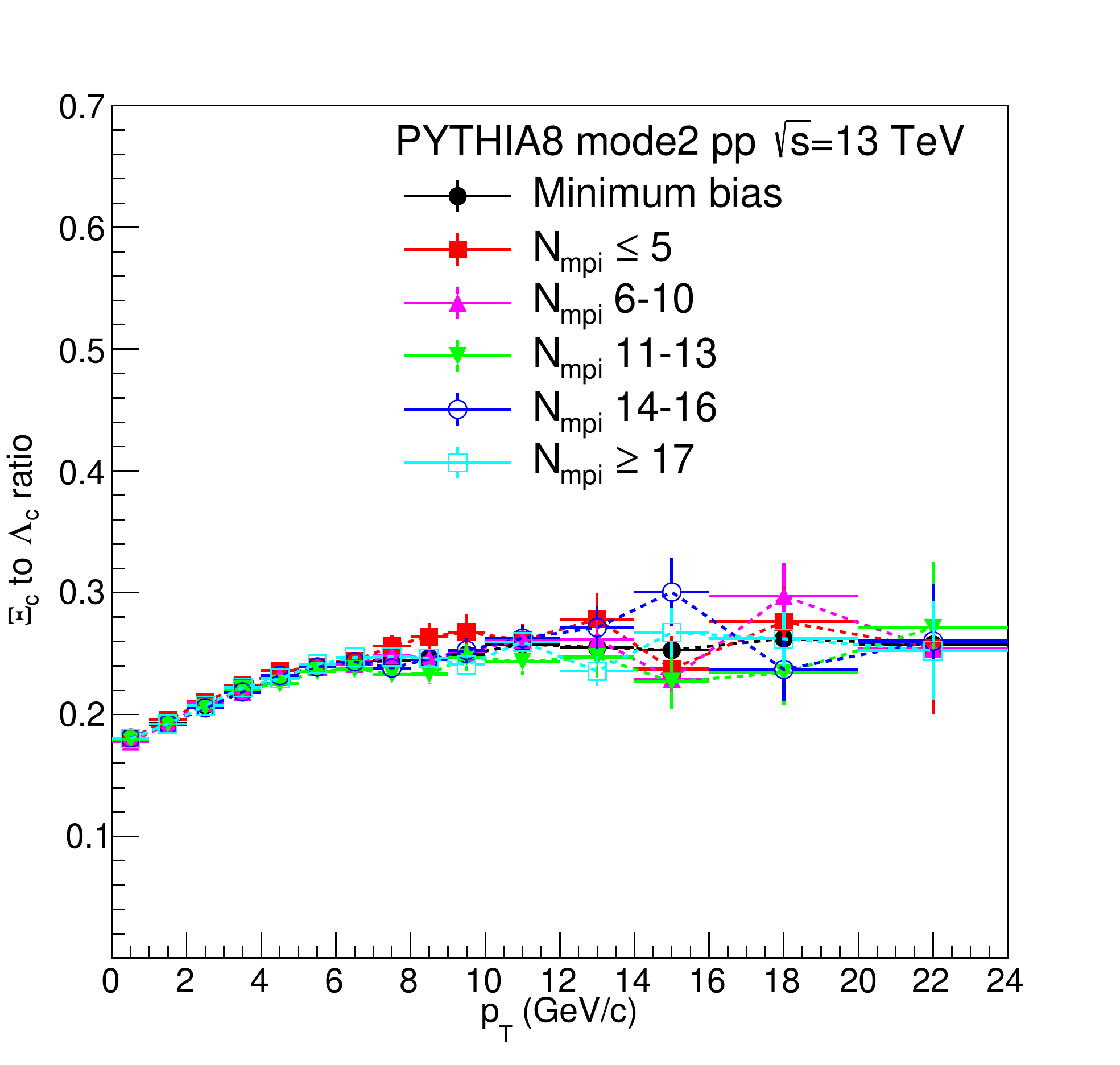}%
\caption{$\Xic/\Dz$ and $\Omc/\Dz$ as well as $\Xic/\Lc$ ratios shown for different \Nmpi classes in the left, center and right panels respectively.}
\label{fig:AllBaryonsVsD0MPI}
\end{figure}

In Fig.~\ref{fig:AllBaryonsVsD0MPI} we show the ratios of the yields of the strange charmed baryon \Xic (left panel) and the double-strange charmed baryon \Omc (central panel) over \Dz as a function of $\pT$ in terms of different \Nmpi classes.
The trends are generally similar to \LcToDz for both \Xic and \Omc. It is to be noted however that, while all ratios fall with \pT, the steepness of the trends are different: while in the range \pT=2 GeV/$c$ to \pT=10 GeV/$c$ the value of \LcToDz falls with a factor of $\approx3$, and $\Sigc/\Dz$ decreases with about a factor of $\approx4$, this decrease is only about a factor of $\approx2$ in the case of \Xic and the \Omc. 
Considering that for the $\Lc$ there is a significant feed-down from $\Xic$ and therefore the result is expected to be a mixture of direct \Lc and those coming from \Xic, this pattern can be attributed to the presence or lack of strange content.

Fig.~\ref{fig:AllBaryonsVsD0MPI} (right) shows the strange-to-non-strange charmed baryon ratio $\Xic / \Lc$ in function of \pT for several \Nmpi classes. Most notable in the figure is that there is no significant \Nmpi ordering, indicating that the event-activity-dependent production is predominantly connected to the charm content and it is not affected by strangeness content. The slight dependence on \pT can be the consequence of the different masses and an overall relative suppression of strangeness production at low \pT.
It is also to be noted that the \Nmpi-dependent enhancement in the strange baryon-to-meson ratio $\Lambda^0/{\rm K}_S^0$ is predicted by PYTHIA 8 alone, albeit to a smaller extent compared to PYTHIA with CR-BLC.

In the following we investigate some distinctive experimental signatures of the enhancement of different charm baryons recapitulated above. To focus on the differences caused by the isospin and strangeness content, we are looking at the baryon-to-baryon ratios. Following the method outlined in Ref.~\cite{Varga:2021jzb} we take the integral of the ratios in the semi-soft (coalescence) regime $2 < \pT < 8$ GeV/$c$.
To characterize the UE and the jettiness of an event, we use a single variable, spherocity, in minimum-bias data, as well as the \RNC and \RT variables in events where a high-momentum trigger hadron is present.
 
\begin{figure}[ht!]
\includegraphics[width=0.33\columnwidth]{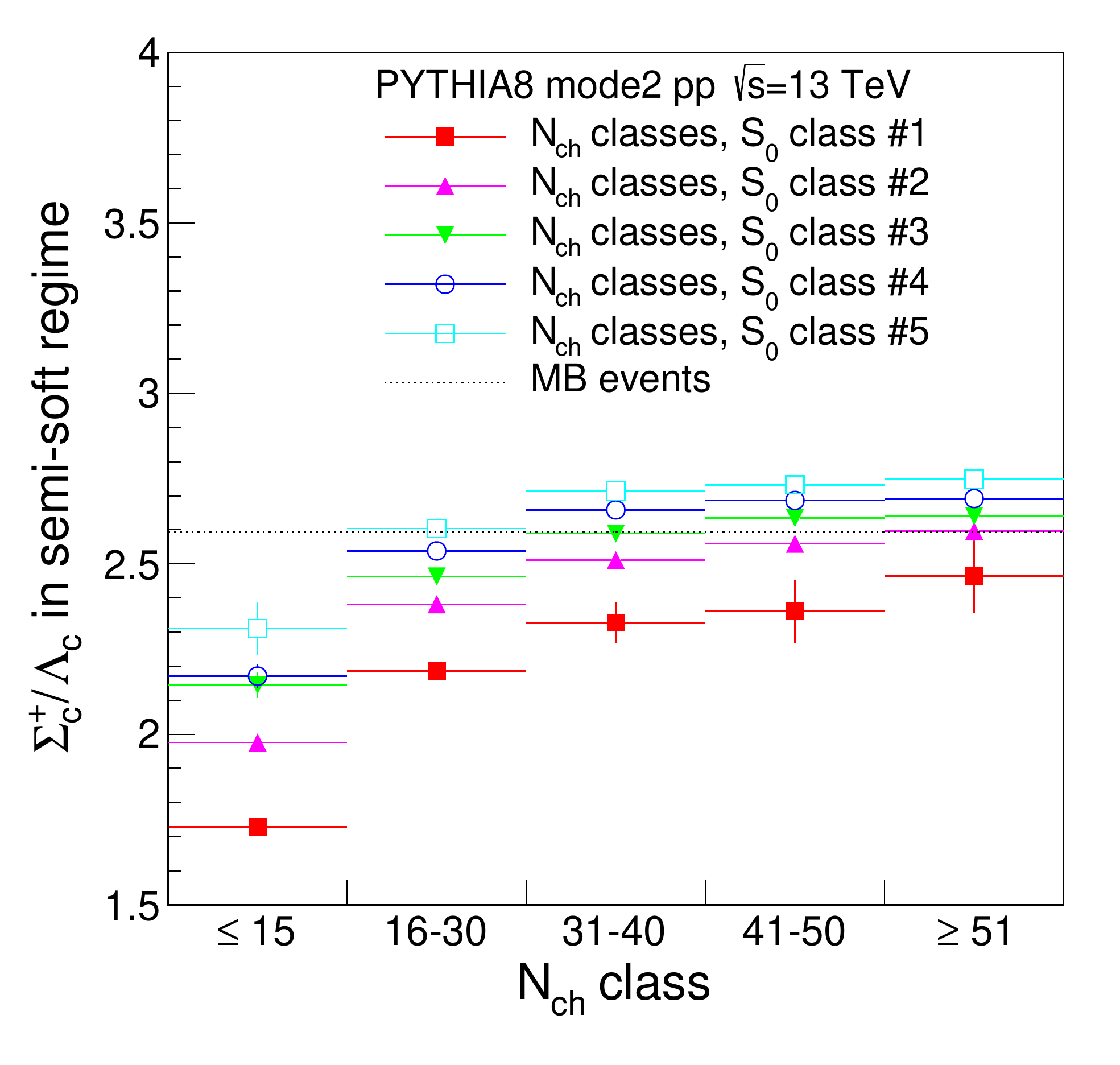}%
\includegraphics[width=0.33\columnwidth]{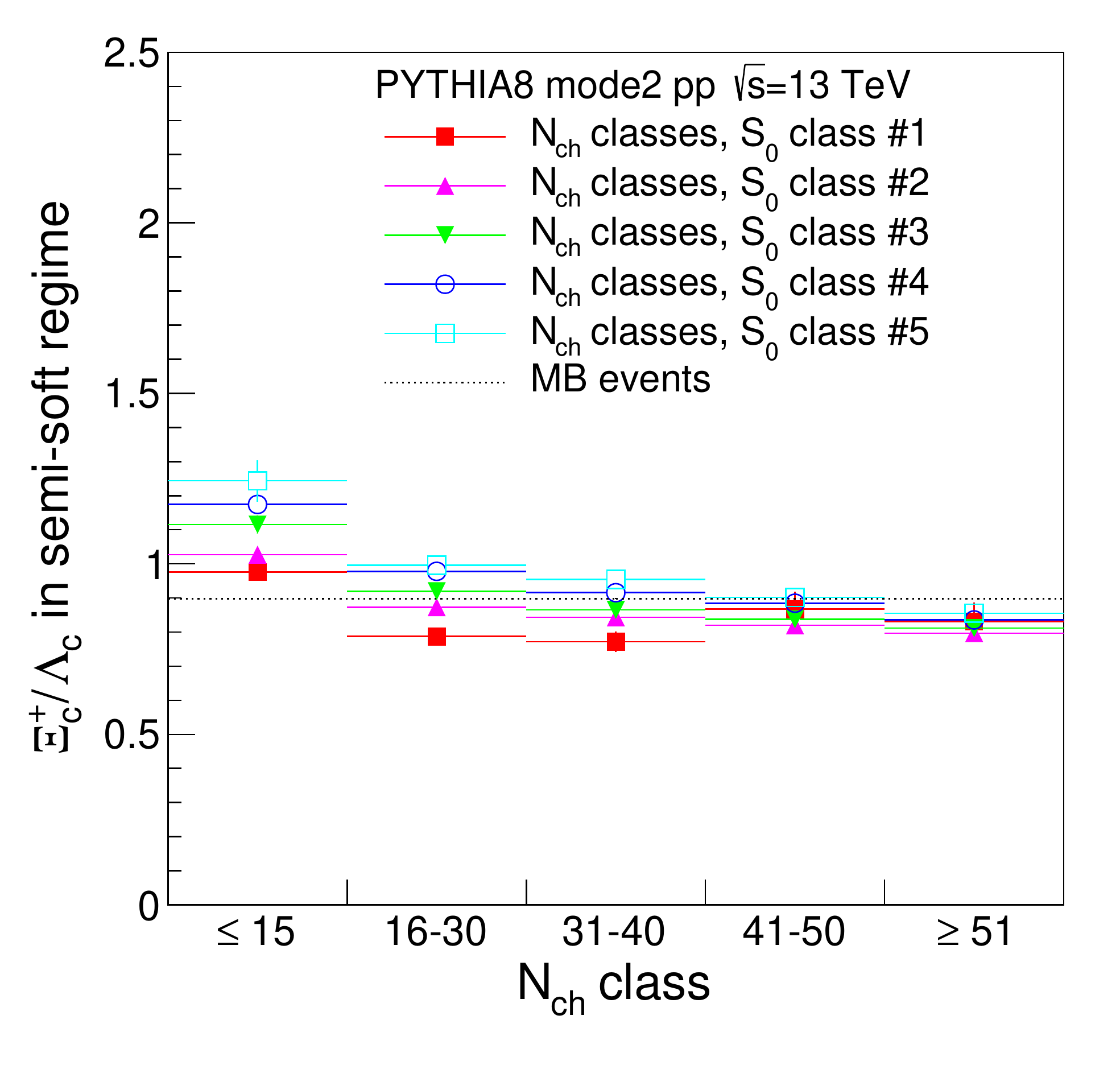}%
\includegraphics[width=0.33\columnwidth]{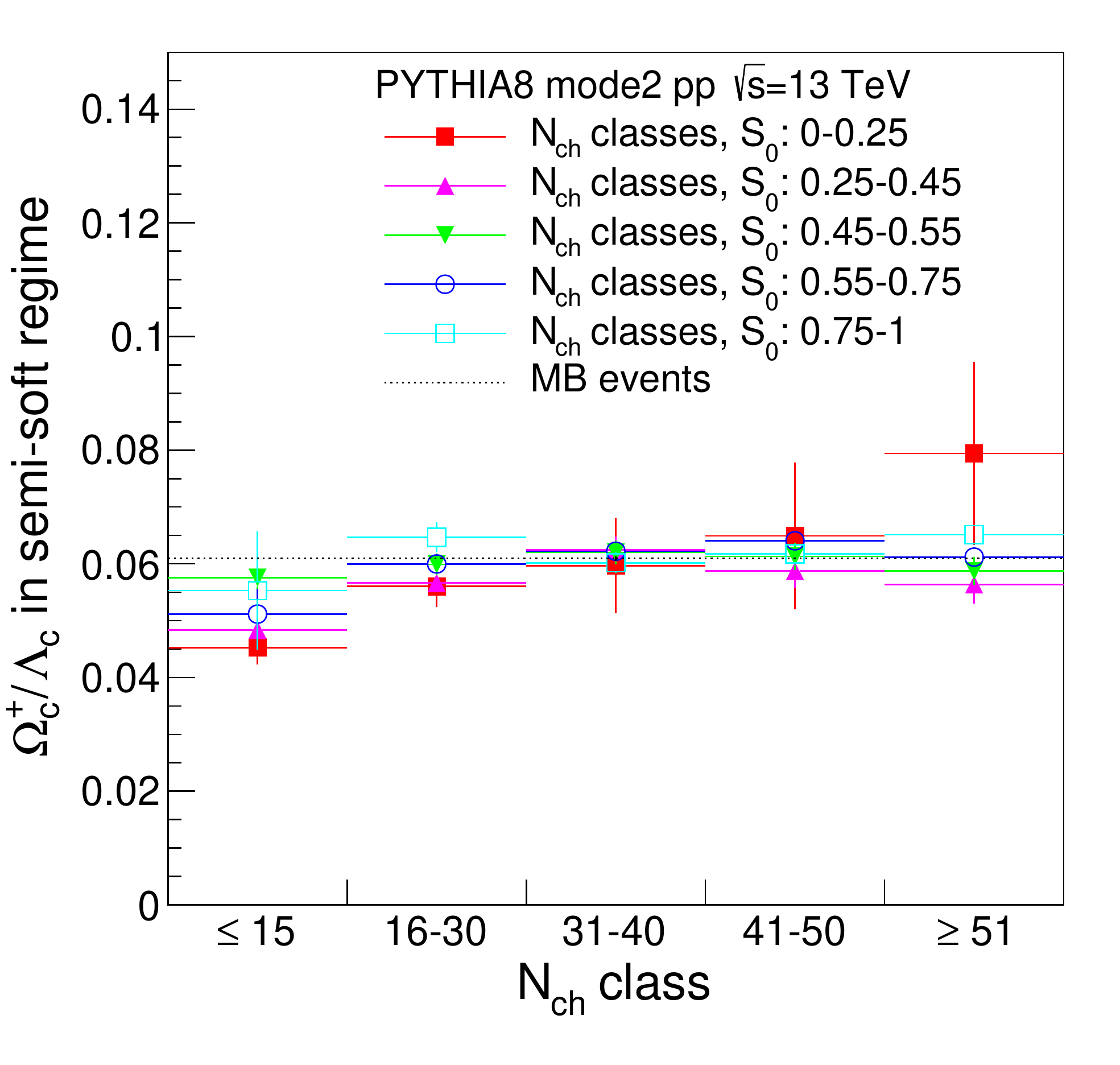}
\caption{Charmed baryon-to-baryon ratios integrated over $2<\pT<8$ GeV/$c$, in fixed \Nch ranges, for different $S_0$ classes (colored curves). $\Sigc/\Lc$ is shown in the left, $\Xic/\Lc$ in the center and $\Omc/\Lc$ in the right panel.}
\label{fig:IntegratedS0Nch}
\end{figure}

In Fig.~\ref{fig:IntegratedS0Nch} we show the $\Sigc/\Lc$ (left), $\Xic/\Lc$ (center) and $\Omc/\Lc$ (right) ratios integrated over the coalescence regime, in fixed \Nch ranges, for different $S_0$ classes. 
Using fixed multiplicity windows reduces the bias from the correlation of jet production with multiplicity.
%
In case of $\Sigc/\Lc$, the ratio consistently depends on $S_0$ in all \Nch classes. This is consistent with the pattern observed in the case of \Nmpi and hints that the enhanced charm-baryon production is sensitive to the isospin. 
On the other hand, strangeness content has only a slight effect in the semi-soft (coalescence) regime. Note that different decay topologies of higher-mass states may have an effect in the lowest \Nch range where the UE is small.
 
\begin{figure}[ht!]
\includegraphics[width=0.33\columnwidth]{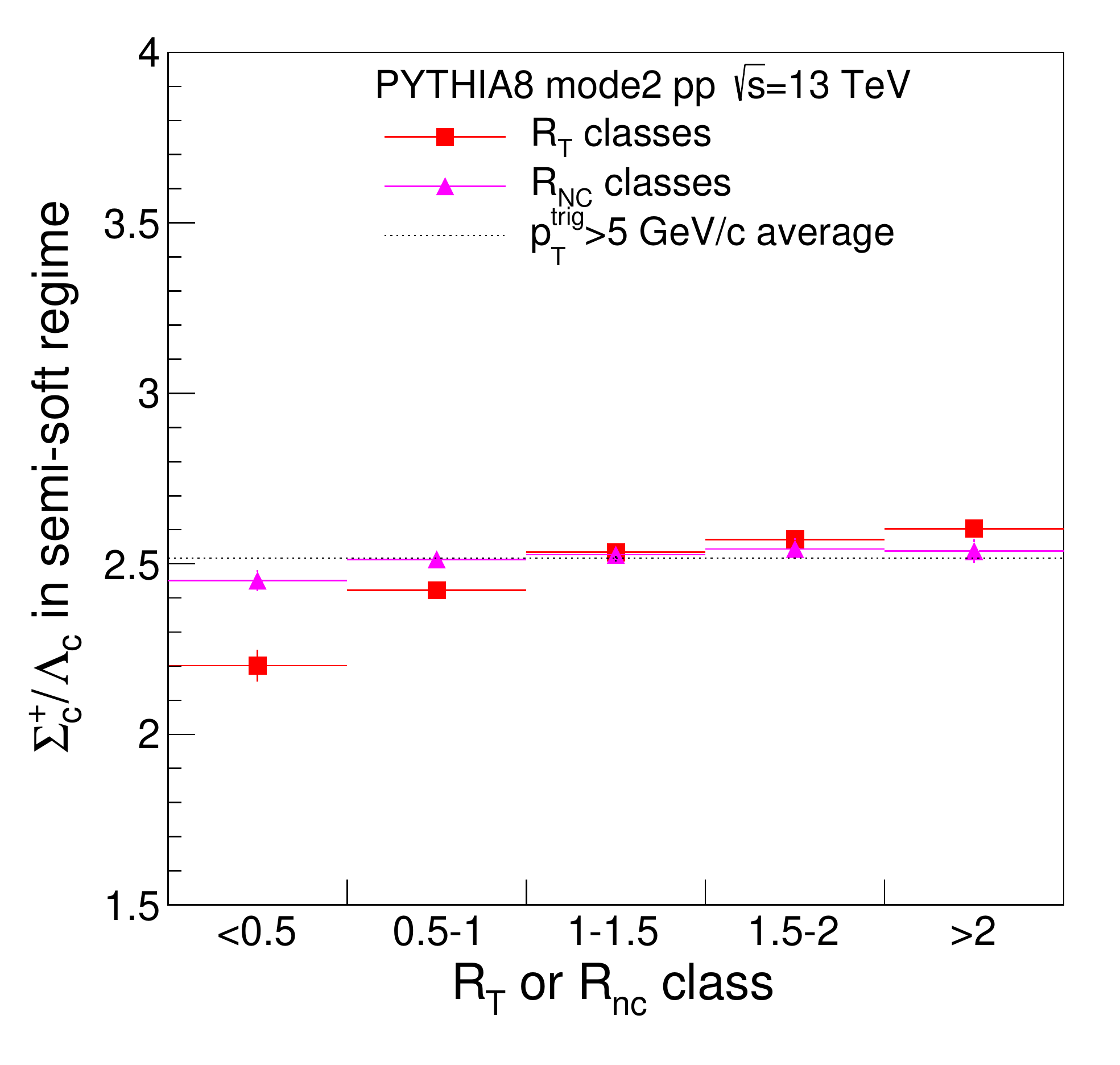}%
\includegraphics[width=0.33\columnwidth]{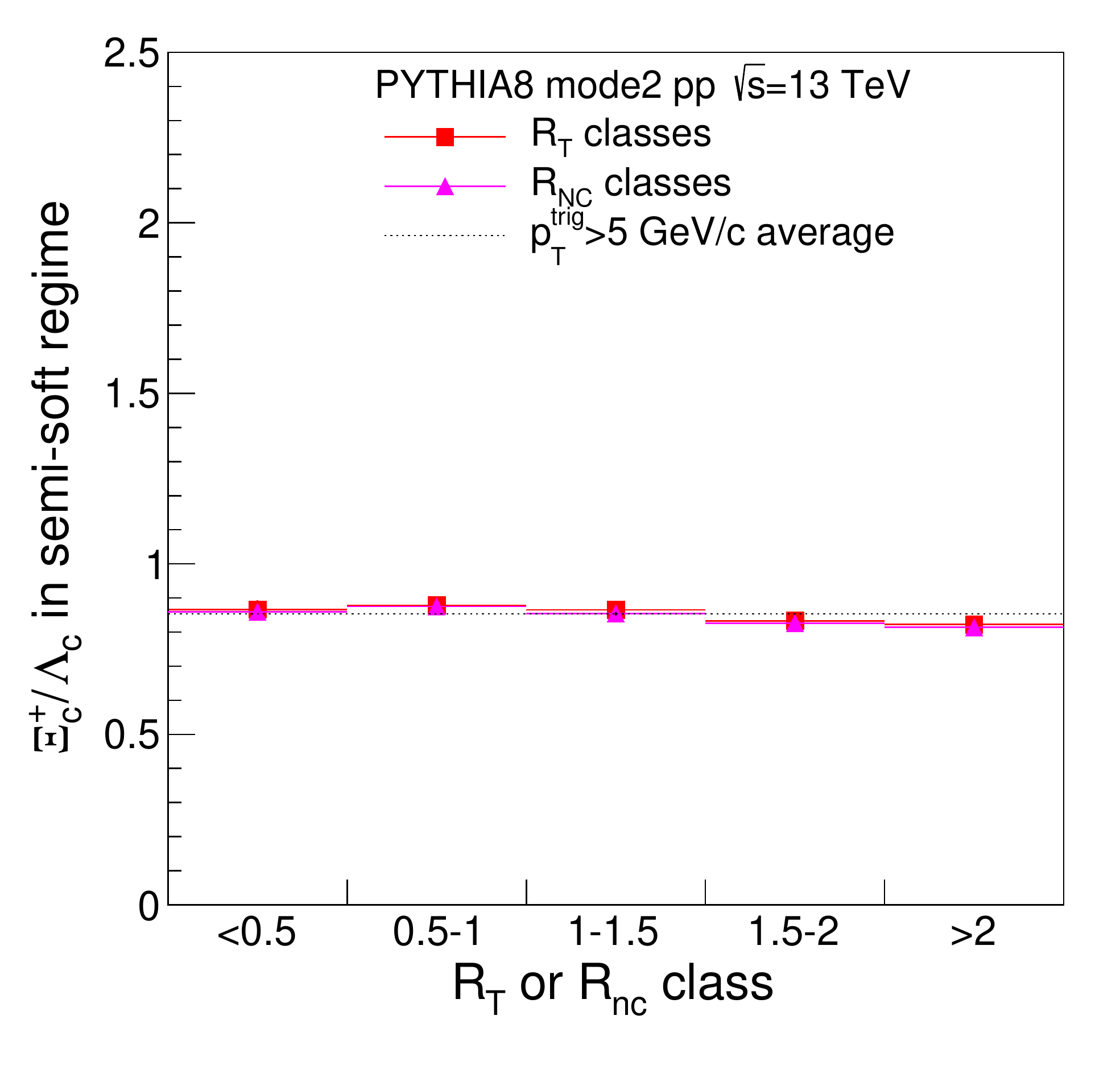}%
\includegraphics[width=0.33\columnwidth]{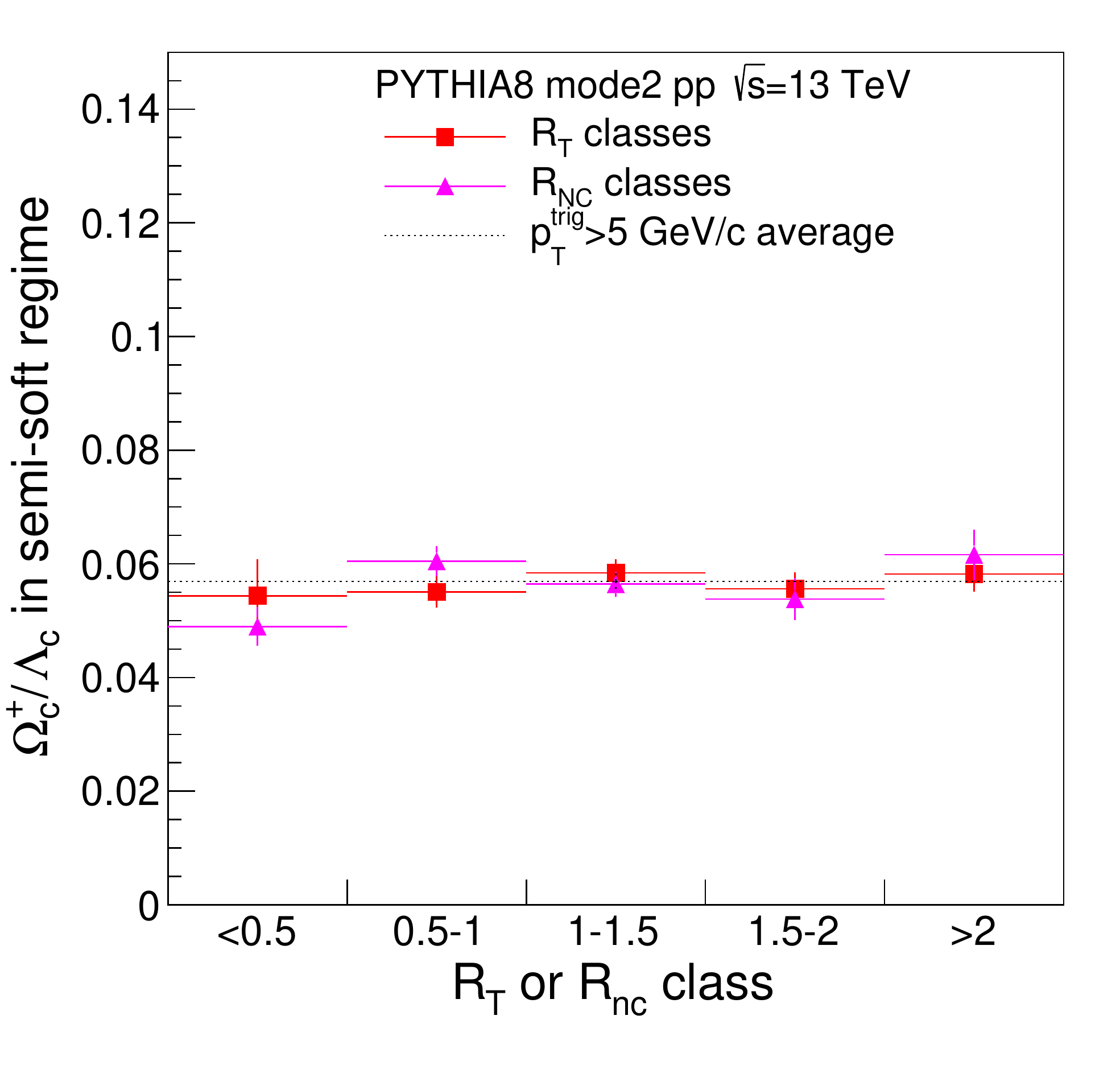}
\caption{Charmed baryon-to-baryon ratios integrated over $2<\pT<8$ GeV/$c$, for different \RT (red) and \RNC (magenta) classes. $\Sigc/\Lc$ is shown in the left, $\Xic/\Lc$ in the center and $\Omc/\Lc$ in the right panel.}
\label{fig:IntegratedRtRnc}
\end{figure}

In Fig.~\ref{fig:IntegratedRtRnc} we plot the integrated $\Sigc/\Lc$ (left), $\Xic/\Lc$ (center) and $\Omc/\Lc$ (right) ratios in the \RT and \RNC classes.
While there is virtually no \RNC dependence for the $\Sigc/\Lc$ ratio, the \RT-dependence is significant. This corroborates the observation that in the PYTHIA 8 CR-BLC model the enhancement is primarily linked to the UE and not to the jet production.
In case of the strange to non-strange charmed baryon ratios, no dependence is observed in either case.

\section{Conclusions}

Enhancement of charmed baryons compared to \Dz and \Lc in pp collisions compared to $e^-e^+$ collisions question the universality of charm fragmentation. We proposed event-activity classifiers which provide great sensitivity to the production mechanism. These observable are directly accessible for measurement in LHC Run 3.

In the model class considering color reconnection beyond leading color approximation,
\Lc is sensitive to the underlying-event activity. While this sensitivity is present both in directly produced as well as decay \Lc baryons, the excess is dominated by the decay contribution at low \pT ($\lesssim 8$ GeV/$c$), and by the direct production at higher \pT ($\gtrsim 16$ GeV/$c$) values. This is not a consequence of decay kinematics.

While strangeness enhancement itself is reproduced by models without color junctions, the enhanced production of charm baryons (either with strange or non-strange content) requires color reconnection beyond leading order. In both the strange and non-strange cases, charm baryon enhancement comes from the underlying event and not from the jet region.
While the isospin-dependent effects seem to be linked to the formation of charm baryons via color junctions, strangeness does not play a strong role in the enhancement of charmed baryons.

Using the proposed observables, the high-precision experimental data from the upcoming LHC Run 3 data taking period will be able to differentiate between mechanisms of strangeness and charm enhancement.

\section*{Acknowledgements}

This work has been supported by the NKFIH grants OTKA FK131979 and K135515, as well as by the 2021-4.1.2-NEMZ\_KI-2022-00007
project.
The authors acknowledge the computational resources provided by the Wigner GPU Laboratory and research infrastructure provided by the Eötvös Loránd Research Network (ELKH).

\vspace{6pt} 

\reftitle{References}


\externalbibliography{yes}
\bibliography{VMVcsbaryon.bib}

\end{document}